\documentclass[preprint,12pt]{elsarticle}
\biboptions{sort&compress}

\usepackage{graphicx}
\usepackage{amssymb}
\usepackage{lineno}
\usepackage{outlines}
\usepackage[dvipsnames]{xcolor} 
\usepackage[normalem]{ulem} 
\journal{Current Opinion in Chemical Engineering}

\begin{document}

\begin{frontmatter}


\title{Modeling hydrodynamic interactions in soft materials with multiparticle collision dynamics}

\author{Michael P. Howard}
\address{McKetta Department of Chemical Engineering, University of Texas at Austin, Austin, TX 78712}
\ead{mphoward@utexas.edu}

\author{Arash Nikoubashman}
\address{Institute of Physics, Johannes Gutenberg University Mainz, Staudingerweg 7, 55128 Mainz, Germany}
\ead{anikouba@uni-mainz.de}

\author{Jeremy C. Palmer}
\address{Department of Chemical and Biomolecular Engineering, University of Houston, Houston, TX 77204}
\ead{jcpalmer@uh.edu}

\begin{abstract}
Multiparticle collision dynamics (MPCD) is a flexible and robust mesoscale computational technique for simulating solvent-mediated hydrodynamic interactions in soft materials.  Here, we provide a critical overview of the MPCD method and summarize its current strengths and limitations.  The capabilities of the method are highlighted by reviewing its recent applications to simulate diverse phenomena, ranging from the flow of complex fluids and thermo-osmotic transport to bacterial swimming and active particle self-assembly. We also discuss outstanding challenges and emerging methodological developments that are expected to greatly expand the applicability of MPCD to other systems of technological importance.
\end{abstract}


\end{frontmatter}


\section{Introduction} \label{S:1}

\par Solvent-mediated hydrodynamic interactions (HI) play an important role in numerous applications, including self-assembly of active particles, mixing of (complex) liquids, and transport in microfluidic devices. Even at equilibrium, HI can have a dramatic effect on dynamic processes like diffusion. Computer simulations are powerful tools for studying these systems, providing microscopic insights into their static and dynamic properties. Simulations can also be used to explore large parameter spaces and to develop predictive theoretical models used in process design. Nonetheless, simulating soft materials in solution often presents a considerable computational challenge. Many soft materials consist of mesoscopic solutes such as colloids or polymer chains that can be orders of magnitude larger than solvent molecules and have correspondingly slower dynamics. Classical molecular dynamics (MD) simulations using an explicit, atomistically detailed solvent are often intractable for these systems because they require large numbers of atoms and many numerical integration steps to probe the length and time scales of interest.

\par Often, fine resolution of the solvent is unnecessary, and only the solvent's effects on the solute need to be described to capture the relevant physics of the system. Such problems are amenable to a mesoscale simulation approach in which the microscopic details of the solvent are simplified by coarse-graining. Various methods have been developed for modeling HI in a coarse-grained fashion, including lattice-based models such as the Lattice-Boltzmann technique \cite{chen:rev:1998, akker:rev:2018}; implicit-solvent models like Brownian dynamics \cite{allen:book:2017} and Stokesian dynamics \cite{brady:rev:1988}; and particle-based models such as dissipative particle dynamics (DPD) \cite{hoogerbrugge:epl:1992, espanol:jcp:2017} and multiparticle collision dynamics (MPCD) \cite{malevanets:jcp:1999}. In this review, we focus on the MPCD method, which has proven to be a useful tool for studying both equilibrium and nonequilibrium behaviors of soft materials. We describe the MPCD algorithm in Section~\ref{S:2}, highlight its recent applications to soft materials in Section~\ref{S:3}, and then conclude by discussing outstanding challenges and future avenues for development. 

\section{Fundamentals} \label{S:2}
\subsection{Algorithm} \label{S:2:algorithm}

\par
\begin{figure}[!htbp]
\centering
\includegraphics[width=\textwidth]{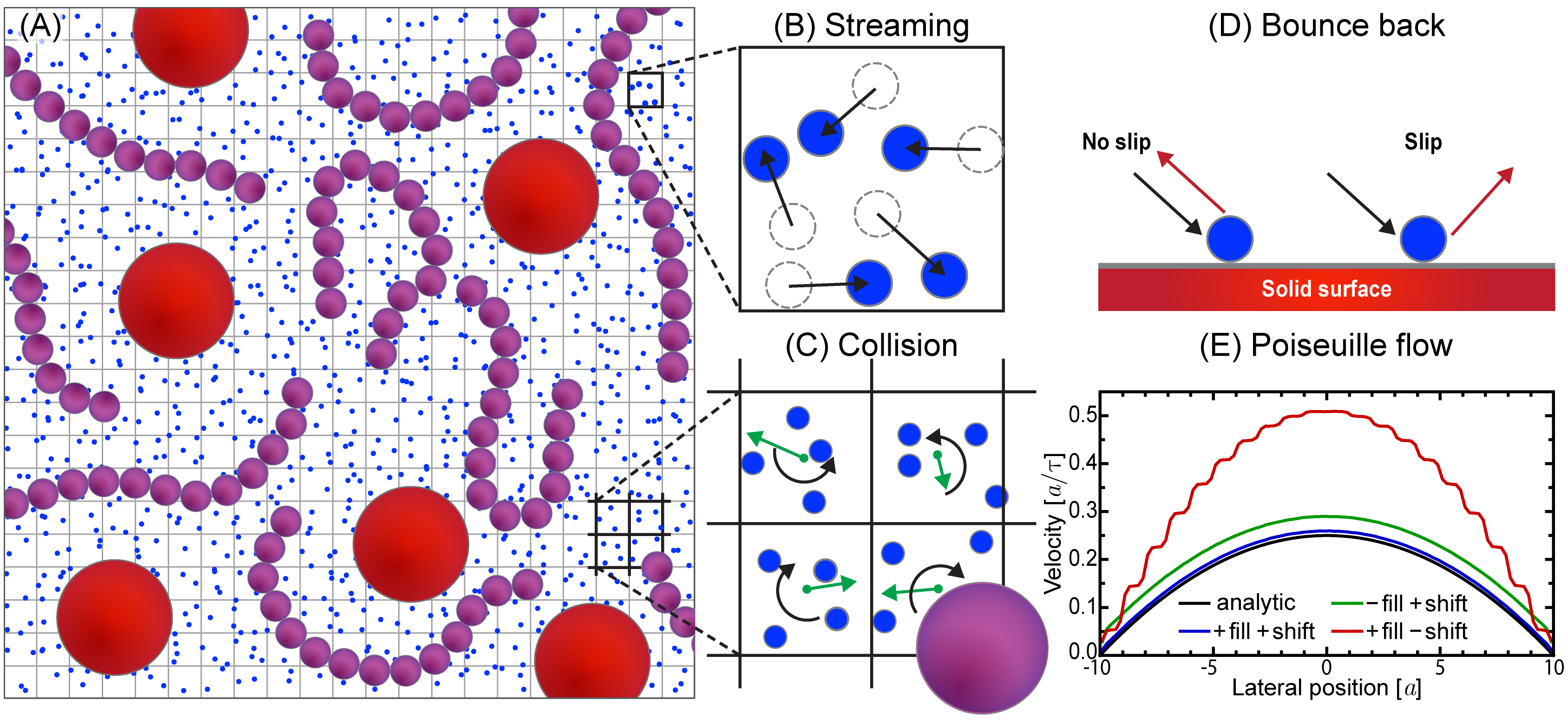}
\caption{(A) Illustration of colloids (red) and polymer chains (purple) embedded in an MPCD solvent (blue). The solutes are propagated using conventional MD and exchange momentum with the MPCD solvent during the streaming and/or collision steps. The MD time step for the solutes $\delta t$ is typically much smaller (a factor of 10 or more) than the time between consecutive multiparticle collisions, $\Delta t$. Hence, MPCD particles are not always updated during intermediate steps of $\delta t$, depending on the solute coupling scheme (Section \ref{S:2:couple}). (B) During the streaming step, solvent particle positions are updated via eq. \ref{eq:stream}. (C) Momentum exchange between solvent particles is simulated during the collision step (e.g., eq. \ref{eq:srd}). Small solute particles such as polymer segments are commonly coupled to the MPCD fluid through the collision step, in which they are treated similarly to solvent particles. (D) Bounce-back schemes are often employed during the MPCD streaming step to couple the MPCD fluid to large solutes (e.g., colloids). The normal component of the velocity relative to the surface is reflected to prevent fluid particles from penetrating the solid surface. Reflecting the tangential component of the velocity gives rise to a no-slip condition at the surface, whereas leaving the tangential component unchanged results in a slip (no stress) boundary condition. These bounce-back rules can also be used to introduce solid boundaries into the system, e.g. channel walls. Inserting planar surfaces with no-slip boundaries and moving them relative to each other generates a shear (Couette) flow. (Alternatively, shear flow can also be generated in MPCD using Lees--Edwards periodic boundary conditions \cite{lees:jpc:1972} or the M\"{u}ller-Plathe reverse nonequilibrium scheme \cite{mueller-plathe:pre:1999}.) Parabolic (Poiseuille) flow can be produced by applying a constant body force to an MPCD solvent between stationary, no-slip parallel plates. As (E) shows, however, the expected parabolic flow profile is only recovered when virtual particles (+fill) are placed inside the solid surfaces and random grid shifts (+shift) are used to ensure Galilean invariance. Significant slip at the walls is observed in simulations without virtual particles (-fill), while discretization artifacts arise without random grid shifts (-shift). \label{fig:1}}
\end{figure}

 The MPCD method was first proposed by Malevanets and Kapral \cite{malevanets:jcp:1999} and is sometimes referred to as stochastic rotation dynamics (SRD) based on their initial variant of the algorithm. The general spirit of MPCD, and other particle-based methods like DPD, is to adopt a computationally inexpensive, coarse grained solvent model that faithfully reproduces the solvent-mediated HI (Fig.~\ref{fig:1}). The MPCD solvent is a fluid of point particles, each having mass $m$, position $\mathbf{r}$, and velocity $\mathbf{v}$. The particle positions evolve according to Newton's equations of motion. In the absence of external forces, the positions can be simply stepped forward over a time interval $\Delta t$,
\begin{equation}
    \mathbf{r}_i(t+\Delta t) = \mathbf{r}_i(t) + \mathbf{v}_i(t) \Delta t,\label{eq:stream}
\end{equation}
where the subscript $i$ denotes the particle index. Equation~\ref{eq:stream} is often referred to as the ``streaming'' step of the MPCD algorithm (Fig.~\ref{fig:1}(B)).

\par After streaming, the solvent undergoes a multiparticle collision that is the core of the MPCD algorithm (Fig.~\ref{fig:1}(C)). The collision can be performed according to various schemes \cite{malevanets:jcp:1999, allahyarov:pre:2002, ihle:epl:2006, noguchi:epl:2007, muhlbauer:prf:2017}, but all share a few common traits. First, they conserve linear momentum to ensure development of hydrodynamic correlations. Second, they are usually spatially localized and simple (e.g., avoiding pairwise force calculations) for computational efficiency. Finally, they are designed so that the MPCD model reproduces as many relevant transport coefficients and properties of the real solvent as possible, especially the kinematic viscosity and diffusivity, which are intimately connected with the propagation of HI.

\par In the original algorithm proposed by Malevanets and Kapral \cite{malevanets:jcp:1999}, commonly referred to as SRD, collisions are handled by sorting MPCD particles into cubic cells with edge length $a$, which sets the spatial resolution of HI \cite{huang:pre:2012}. The center-of-mass velocity of each cell, $\mathbf{u} = \sum m_i \mathbf{v}_i / \sum m_i$, is determined. All particles in the same cell then undergo a stochastic collision,
\begin{equation}
    \mathbf{v}_i(t+\Delta t) = \mathbf{u}(t) + \mathbf{R} \cdot \left(\mathbf{v}_i(t) - \mathbf{u}(t) \right) ,
    \label{eq:srd}
\end{equation}
in which the particle velocities relative to $\mathbf{u}$ are rotated by a norm-conserving rotation matrix $\mathbf{R}$ chosen randomly and independently for each cell. Most commonly, $\mathbf{R}$ rotates the velocities in a cell by a fixed angle $\alpha$ around an axis randomly selected from the unit sphere \cite{allahyarov:pre:2002}. Equation~\ref{eq:srd} conserves both linear momentum and energy within a cell.

\par In principle, only eqs.~\ref{eq:stream} and \ref{eq:srd} are needed to perform an MPCD simulation. Malevanets and Kapral demonstrated that this algorithm obeys the H theorem, meaning that the particle velocities relax to a Maxwell--Boltzmann distribution, and more importantly, that the algorithm yields the hydrodynamic equations for compressible flow \cite{malevanets:jcp:1999}. Nonetheless, additional modifications to the MPCD algorithm are often needed. For instance, discretization of the MPCD particles into collision cells can lead to the loss of Galilean invariance when the particle mean-free path is smaller than the collision cell, resulting in spurious correlations between particle collisions. This issue is typically remedied by randomly shifting the origin of the MPCD cell grid before each collision step to restore Galilean invariance \cite{ihle:pre:2001}.

\par As another example, eq.~\ref{eq:srd} conserves linear momentum but not angular momentum \cite{gotze:pre:2007}. Angular momentum conservation can be restored for eq.~\ref{eq:srd} \cite{noguchi:epl:2007}; the cost, however, is loss of energy conservation. The MPCD particles must then be coupled to a thermostat to maintain constant temperature, which is also required for nonequilibrium simulations. The most common thermostat for the SRD rule rescales $\mathbf{v}$ relative to $\mathbf{u}$ for each cell by a factor that enforces the correct temperature and ensemble \cite{huang:pre:2015}. Alternatively, temperature control can be incorporated directly into the collision rule, as in the Andersen thermostat (AT) scheme \cite{allahyarov:pre:2002}. Additional technical details of the basic MPCD algorithm and its many variants can be found in Refs.\ \cite{kapral:advchem:2008, gompper:advcomp:2009}.

\subsection{Mapping} \label{S:2:mapping} 
\par One beneficial feature of MPCD is that the solvent transport coefficients can be estimated analytically \cite{ihle:pre:2003, padding:pre:2006}. For the commonly used SRD parameters of rotation angle $\alpha = 130^\circ$, collision time $\Delta t = 0.1\,\tau$, and particle density $\rho = 5\,m/a^3$, the analytic expressions predict solvent shear viscosity $\mu = 3.96\,k_{\rm B}T \tau/a^3$ and self-diffusivity $D = 0.064\,a^2/\tau$, where $\tau = \sqrt{m a^2/k_{\rm B} T}$ is the MPCD unit of time, $k_{\rm B}$ is Boltzmann's constant, and $T$ is temperature. Analytic estimates of $\mu$ are typically in good agreement with direct calculations from simulation, whereas estimates for $D$ are smaller than the measured values due to hydrodynamic correlations not accounted for in the analytic expressions \cite{ripoll:pre:2005}.

\par It is tempting to map the MPCD solvent model onto a real liquid like water. We set $T = 298\,{\rm K}$ and the cell size $a = 3\,{\rm \mbox{\AA}}$ to be comparable to the molecular diameter of water. If the MPCD solvent density is matched to that of water ($\rho = 997\,{\rm kg}/{\rm m}^3$) through $m$, the viscosity of the MPCD solvent is $0.21\,{\rm mPa}\,{\rm s}$, which is lower than the viscosity of water ($0.89\,{\rm mPa}\,{\rm s}$) \cite{nistwebbook}. Repeating this mapping for hexane, another common solvent used in experiment, which has as a lower density ($\rho = 655\,{\rm kg}/{\rm m}^3$) and viscosity ($0.30\,{\rm mPa}\,{\rm s}$) than water \cite{nistwebbook}, gives a model viscosity ($0.17\,{\rm mPa}\,{\rm s}$) that is closer to the experimental value. The corresponding MPCD time step under both mappings is roughly 30 fs. This value is significantly larger than the ca.\ 1 -- 2 fs time steps used in atomistic MD simulations, thereby allowing longer effective time scales to be probed.

\par This mapping has several drawbacks. First, the viscosities are not a perfect match, and the length scale $a$ (and time step) would need to be drastically reduced to obtain better agreement. Second, even taking $a = 3\,{\rm \mbox{\AA}}$ may lead to prohibitively large MPCD simulations when the solute is big (tens of nanometers). An alternative mapping based on the solute, which sacrifices properties of the fluid, may be necessary in these cases \cite{padding:pre:2006}. Finally, the solvent diffusivity maps to $D \approx 16.8\,\mu{\rm m}^2/{\rm ms}$ for water, which is an order of magnitude larger than experimentally measured \cite{wang:jpc:1965}. This mismatch is inherent to the chosen SRD parameters and mapping, as is apparent from the Schmidt number, ${\rm Sc} = \mu/\rho D \approx 12$. Although this value of Sc is usually considered liquid-like for MPCD \cite{padding:pre:2006,ripoll:pre:2005}, it is an order of magnitude smaller than the values for typical solvents such as water.  The implications of the MPCD solvent (and other models like DPD and the Lennard-Jones fluid \cite{bolintineanu:cpm:2014}) having low Schmidt number are not fully understood and require further investigation.

\subsection{Solute and boundary coupling} \label{S:2:couple}
\par A notable advantage of MPCD compared to other techniques for modeling HI in soft materials is its compatibility with a variety of complex solutes and boundaries \cite{bolintineanu:cpm:2014}. The MPCD solvent has been coupled to solutes such as macromolecules \cite{malevanets:epl:2000}, rigid colloids \cite{malevanets:jcp:2000, inoue:jstat:2002, padding:jpcm:2005, bolintineanu:cpm:2014}, and membranes \cite{noguchi:pre:2005}.  These systems are typically simulated using a hybrid MD--MPCD scheme in which the solute is propagated using conventional MD techniques while also exchanging momentum with the MPCD solvent during the streaming and/or collision steps. 

\par Macromolecules like polymers are coupled to the MPCD solvent during the collision step \cite{malevanets:epl:2000} as ``heavy'' particles \cite{malevanets:epl:2000} (Fig.~\ref{fig:1}(C)). The resulting dynamics of a linear polymer chain are consistent with the Zimm model \cite{mussawisade:jcp:2005}, which includes pre-averaged HI. Slip or no-slip solid boundaries are incorporated during the streaming step using specular reflection (``bounce-back'') rules \cite{lamura:epl:2001} (Fig.~\ref{fig:1}(D)) or by randomly redrawing the velocities of reflected particles \cite{inoue:jstat:2002, padding:jpcm:2005, whitmer:jpcm:2010, bolintineanu:cpm:2014}. Reflection alone is not sufficient to guarantee that there is no slip at a surface \cite{lamura:epl:2001} because collision cells sliced by the boundary are effectively underfilled with solvent; ``virtual'' solvent particles must be introduced inside the solid surface during the collision step \cite{lamura:epl:2001} (Fig.~\ref{fig:1}(E)). Biasing the velocity of these particles helps enforce the no-slip condition in flow simulations \cite{bolintineanu:pre:2012}.

\par Possibly the most difficult solute to reliably couple to the MPCD solvent is a moving solid boundary such as a spherical colloid \cite{malevanets:jcp:2000, inoue:jstat:2002, padding:jpcm:2005, bolintineanu:cpm:2014} or a deformable membrane \cite{noguchi:pre:2005}. Malevanets and Kapral used short-ranged pair interactions between solvent and solute \cite{malevanets:jcp:2000}, giving slip boundary conditions and requiring costly MD steps to propagate the solvent.  Momentum exchange between the solute and solvent can alternatively be performed using bounce-back rules during streaming \cite{malevanets:jcp:1999, lamura:epl:2001, whitmer:jpcm:2010, bolintineanu:pre:2012}. Inoue et al. proposed another scheme \cite{inoue:jstat:2002} where the solvent is stochastically reflected from the solute surface \cite{padding:jpcm:2005, bolintineanu:cpm:2014}. Virtual particles may be needed to achieve no-slip conditions \cite{lamura:epl:2001}, involving an additional transfer of momentum to the solute \cite{whitmer:jpcm:2010}. Poblete et al. proposed an alternative methodology where the colloid surface is represented by a penetrable set of point particles connected by springs that are included in the collision step \cite{poblete:pre:2014}.

\par It remains an open question as to which coupling scheme is optimal for a given problem. Padding and Louis identified spurious depletion forces between colloids when their interactions allowed overlap of volume excluded to the MPCD solvent by the coupling scheme \cite{padding:pre:2006}. Bolintineanu et al. found that the early- and long-time diffusion coefficients of colloids in suspension depended sensitively on the coupling scheme and collision rule \cite{bolintineanu:cpm:2014}. Cerbelaud et al. measured the shear viscosity of colloidal suspensions and found that the different coupling schemes were largely equivalent in the dilute regime, but only stochastic bounce-back coupling reproduced the correct dynamics at higher volume fractions \cite{cerbelaud:sm:2017}. Recent simulations by Shakeri et al.~\cite{shakeri:pf:2018} challenge whether MPCD with bounce-back rules is able to produce two-particle HI in the Stokes flow limit, but Poblete et al.~\cite{poblete:pre:2014} were able to reasonably reproduce the Stokes flow around a colloid with their scheme. Further study is required to resolve these outstanding issues and discrepancies. At present, we advocate carefully characterizing the colloid dynamics to identify an appropriate coupling scheme for the problem of interest. 

\subsection{Computational performance}
\par The MPCD particle density (usually 5 or 10 particles per $a^3$) is higher than typical atomistic MD models (roughly 3 atoms per $a^3$ for water using the mapping in Section \ref{S:2:mapping}). Hence, MPCD requires more particles than MD to simulate the same system volume. The comparatively simple MPCD solvent interactions and equations of motions (eq.~\ref{eq:stream}) are less computationally demanding, however, and are readily amenable to parallelization to obtain good performance even with large numbers of particles.  Parallel implementations of MPCD have been developed for traditional CPUs \cite{petersen:jcp:2010, debuyl:jopenressoftw:2017}, but large simulations may require hundreds of CPU cores. Recently, the massive parallelism of graphics processing units (GPUs) has been leveraged to accelerate MPCD \cite{westphal:cpc:2014, howard:cpc:2018}. Howard et al.'s open-source implementation of MPCD delivered the performance of nearly 3 CPU nodes (48 cores) using only a single GPU and scaled to run on hundreds of GPUs, enabling simulation volumes as large as $(400\,a)^3$ \cite{howard:cpc:2018}. Moreover, the MPCD simulations of a polymer solution on a GPU were at most a factor of 1.4 slower than a Langevin dynamics simulations without HI, demonstrating the feasibility of routinely performing large-scale MPCD simulations \cite{howard:cpc:2018}. Massively-parallel, open-source implementations of MPCD can be found in current versions of the HOOMD-blue (2.5.0) and LAMMPS (12 Dec 2018) simulation packages.

\section{Applications} \label{S:3}
\subsection{Complex fluids}
\par The MPCD solvent described in Section~\ref{S:2} is a Newtonian fluid, i.e., the shear viscosity $\mu$ is independent of the shear rate. Many biologically and technologically relevant fluids, however, are non-Newtonian complex fluids. For instance, polymer solutions exhibit a distinct viscoelastic response to shear due to the elastic restoring forces counteracting the shear-induced deformation of the polymers. MPCD is well suited to study the dynamics of such complex fluids.

\par Viscoelasticity has been incorporated in MPCD by bonding pairs of solvent particles with harmonic springs to form dumbbells, resulting in a Maxwell fluid \cite{tao:jcp:2008}. The MPCD streaming step is straightforward to implement for this model, but it does not reproduce certain nonequilibrium behaviors like shear-thinning due to the possibility of infinite bond extension. Ji et al. proposed to resolve this issue using finitely-extensible bonds \cite{ji:jcp:2011}, but this modification complicates the streaming step. Kowalik and Winkler instead constrained the bond length with a shear-rate-dependent spring constant \cite{kowalik:jcp:2013}. Their model has a convenient streaming equation, reproduces the expected shear-thinning behavior, and yields a nonzero second normal stress coefficient.

\par Alternatively, viscoelasticity can be modeled by dispersing polymers in the MPCD solvent using the scheme in Section~\ref{S:2:couple}. The equilibrium and nonequilibrium dynamics of flexible polymer chains obtained for dilute \cite{mussawisade:jcp:2005} and semidilute  \cite{huang:mm:2010} solutions were in excellent agreement with predictions of the Zimm model. Solutions of semiflexible polymers \cite{nikoubashman:jcp:2016, nikoubashman:mm:2017} exhibited a significant increase in shear viscosity (and slowdown of chain dynamics) compared to the flexible polymers. Similar models have been used to study the equilibrium dynamics of other macromolecules, including star polymers \cite{ripoll:prl:2006, singh:jcp:2014}, ring polymers \cite{hegde:jcp:2011, liebetreu:ml:2018}, and microgels \cite{ghavami:ml:2017}.

\par
\begin{figure}[!htbp]
\centering
\includegraphics[width=\textwidth]{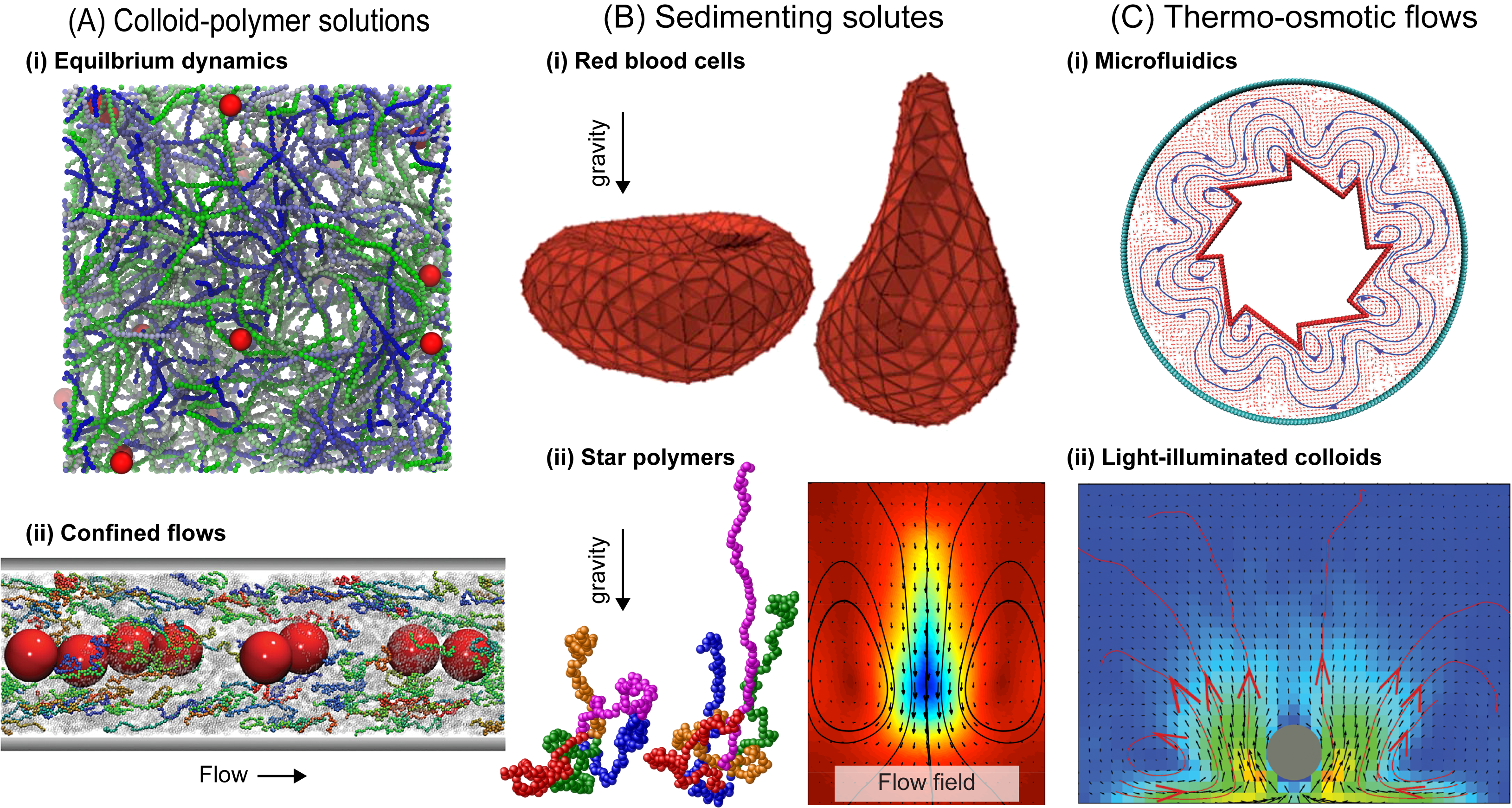} \caption{(A) (i) Colloids suspended in a solution of semiflexible polymers. Reproduced from Ref.\ \cite{chen:sm:2018} with permission from The Royal Society of Chemistry.  (ii) Colloids in dilute solution of flexible polymers flowing through a slit channel. Viscoelastic forces exerted by the polymers push the colloids towards the channel centerline. Reprinted from Ref.\ \cite{howard:jcp:2015}, with the permission of AIP Publishing. (B) Sedimentation of (i) an elastic network model of red blood cells \cite{peltomaeki:sm:2013} and (ii) star polymers under an applied gravitational field \cite{singh:jcp:2018}. (i) Cells with stiff membranes in a weak field adopt a parachute shape (left), whereas those with flexible membranes develop a teardrop-like morphology at high field strength (right). Republished with permission of The Royal Society of Chemistry, from Ref.\ \cite{peltomaeki:sm:2013}; permission conveyed through Copyright Clearance Center, Inc. (ii) (left, middle) Increasing field strength enhances deformation of star polymers. (right) Flow field in the laboratory frame of a sedimenting star polymer. Reprinted from Ref.\ \cite{singh:jcp:2018}, with the permission of AIP Publishing. (C) (i) Flow field and streamlines generated in a microfluidic device consisting of a heated gear-shaped wall circumscribed by a circular wall \cite{yang:sm:2016}. The higher temperature of the gear-shaped wall induces a thermo-osmotic flow that stirs the confined fluid. (ii) Flow field generated when a light-adsorbing colloid (gray circle) is heated under uniform illumination near a thermophobic wall \cite{lou:sm:2018}. Panels (i) and (ii) in (C) were republished with permission of The Royal Society of Chemistry, from Refs. \cite{yang:sm:2016} and \cite{lou:sm:2018}, respectively; permission conveyed through Copyright Clearance Center, Inc.\label{fig:2}} 
\end{figure}

 MPCD has also been applied to study the equilibrium dynamics of colloidal suspensions in Newtonian and non-Newtonian fluids. Several studies have investigated the shear viscosity and diffusivity of hard-sphere colloidal suspensions at volume fractions ranging from the dilute limit to $40\,\%$ \cite{bolintineanu:cpm:2014, cerbelaud:sm:2017}. Recently, the diffusion of colloidal particles in semidilute polymer solutions was studied in the regime where the colloid diameter is comparable to the polymer mesh size \cite{chen:sm:2017, chen:mm:2018, chen:sm:2018} (Fig.~\ref{fig:2}(A)). These simulations revealed an intricate coupling between the colloid motion and the segmental relaxation and center-of-mass motion of the polymers, leading to subdiffusive dynamics for both species at intermediate timescales.

\par In addition to equilibrium dynamics, MPCD is well suited to study the nonequilibrium behaviors of complex fluids. Sedimenting hard-sphere colloids were found to exhibit a hydrodynamic (Rayleigh--Taylor-like) instability \cite{wysocki:faraday:2010}. Sedimenting red blood cells \cite{peltomaeki:sm:2013} were shown to assume different shapes depending on the gravitational strength and the elastic moduli of their membranes (Fig.~\ref{fig:2}(B)), but their terminal velocities were mostly independent of the cell shape. For sedimenting dilute star polymers \cite{singh:jcp:2018}, their shape deformation became more pronounced at increased gravitational strength (Fig.~\ref{fig:2}(B)).

\par MPCD has been used extensively to study complex fluids under flow in microfluidic channels, providing insight into several surprising phenomena. Prohm et al. simulated the inertial focusing of spherical colloids in a tube \cite{prohm:epje:2012} and found that the colloids formed an annulus around the centerline, an analog of the macroscopic Segr{\'e}-Silberberg effect \cite{segre:nat:1961}. The annular distribution was smeared by thermal fluctuations, but sharpened for larger colloids and faster flows. Rigid colloids in dilute polymer solutions instead focused onto the channel center under specific conditions \cite{howard:jcp:2015} (Fig.~\ref{fig:2}(A)). Dense colloidal suspensions in pressure-driven flow were found to form periodic velocity and density pulse trains, originating from jamming and release of the colloids \cite{kanehl:prl:2017}. Self-assembled micelles of amphiphilic Janus colloids grew in size and become more symmetric under intermediate shear flows before eventually breaking up under larger shear rates \cite{nikoubashman:sm:2017}.

\par Recently, the influence of thermal gradients on flow behavior was also investigated with MPCD. Yang and Ripoll showed how microchannels with ratcheted walls can be used as effective microfluidic pumps without moving parts by locally heating the boundaries \cite{yang:sm:2016}. The temperature gradient induces thermo-osomotic flows (Fig.~\ref{fig:2}(C)) that could be exploited to create complex flow patterns \cite{yang:sm:2016} or rotate microgears \cite{yang:sm:2014}. Burelbach et al. studied the forces acting on a stationary colloid inside a temperature gradient \cite{burelbach:sm:2018}. They found that the thermophoretic forces were proportional to the temperature gradient, in agreement with theoretical predictions, and that the magnitude of this force depended on the hydrodynamic boundary conditions of the colloid. Thermo-osmotic flow from a light-adsorbing colloid led to long-ranged interactions that affected the colloid's dynamics near a wall \cite{lou:sm:2018} (Fig.~\ref{fig:2}(C)). In addition, the motion of an optically trapped colloid in a light-adsorbing fluid was studied, with the resulting thermo-osmotic flow exerting a torque on the trapped particle.

\subsection{Active soft matter}
\par An emerging area of application of MPCD is modeling the hydrodynamics of ``active'' systems.  Unlike the examples above, where solute dynamics arise from equilibrium thermal fluctuations or external perturbations, active particles self-propel through solutions by converting available energy into work. Motile microorganisms such as protozoa and bacteria are the canonical biological example of active particles. They use appendages known as flagella to swim through fluids \cite{Conrad2018}. Because of their small size and slow swimming speeds, their HI are described by the linear, inertialess Stokes equations. In this regime, swimming motions that are symmetric upon time reversal result in zero net displacement and hence cannot be used to move through the fluid.  Consequently, flagellated microorganisms evolved to swim using motions that break time-reversal symmetry \cite{Conrad2018}. 

\par
\begin{figure}[!htbp]
\centering
\includegraphics[width=\textwidth]{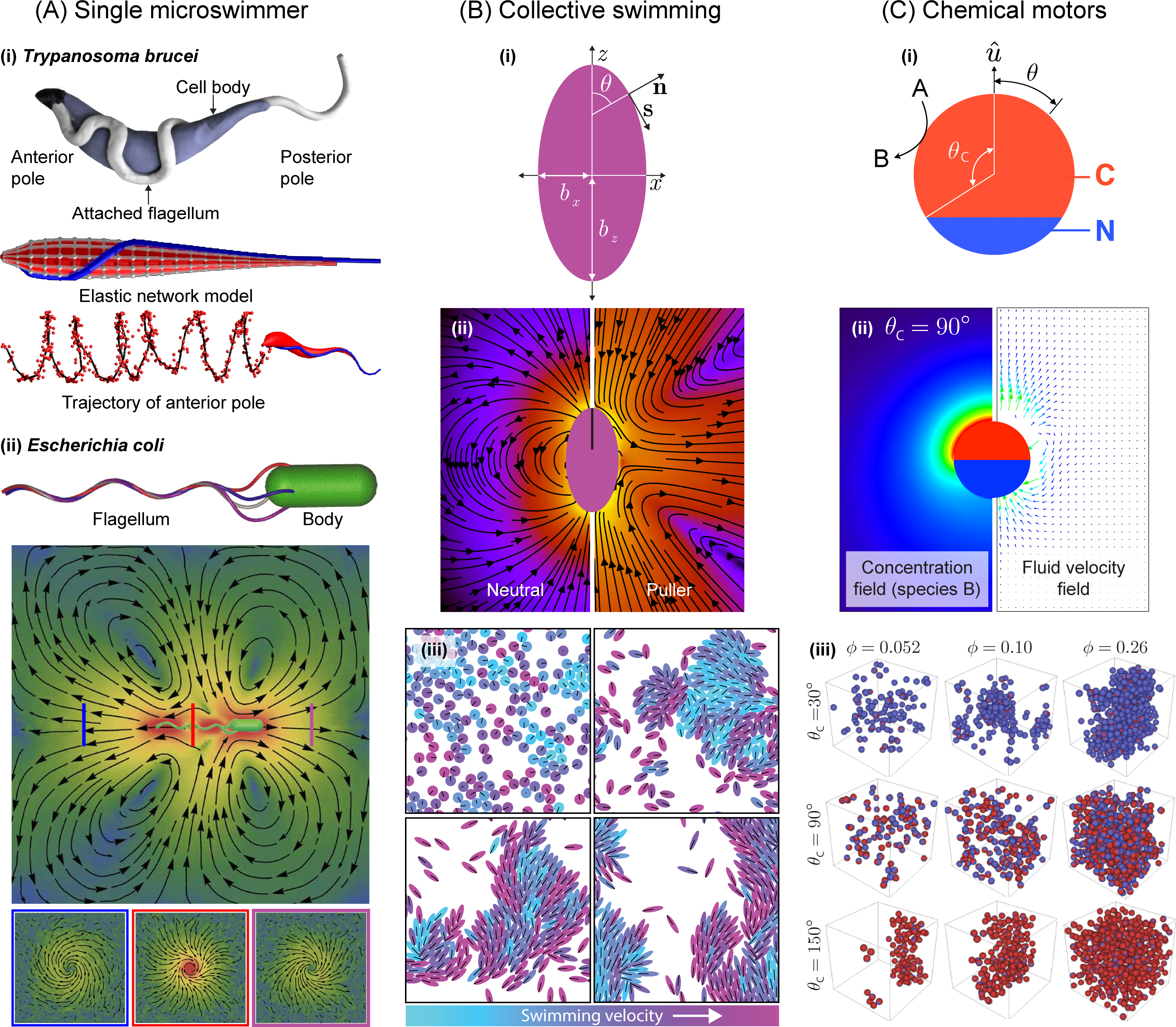}
\vspace{-25pt}
\caption{(A) (i) (top) Anatomy and (middle) elastic network model  of \textit{Trypanosoma brucei}. (bottom) The model captures the cork-screw shaped forward swimming mode observed in experiment. Top and bottom/middle panels were adapted from Refs. \cite{Heddergott2012} and \cite{Alizadehrad2015}, respectively, under a CC BY 4.0 licence [https://creativecommons.org/licenses/by-sa/4.0/]. (ii) (top) Model of \textit{Escherichia coli} with four helical flagella. (middle) The flagellar bundle propels the bacterium and induces a flow field. (bottom) Cross sections of the flow field at positions indicated by vertical lines. Adapted from Ref.\ \cite{Hu2015} under a CC BY 3.0 license. Published by The Royal Society of Chemistry. (B) (i) Prolate spheroidal squirmer model. Propulsion along $z$ is achieved by applying a slip velocity at the squirmer's surface in the direction of tangent $\mathbf{s}$. The slip velocity specifies the squirmer type: neutral swimmer, puller, or pusher. (ii) Flow fields generated by neutral swimmers and pullers; flow near pushers is similar to pullers, but with arrow directions reversed. (iii) Collective behavior of neutral swimmers with different aspect ratios. Panels (i,ii) and (iii) were adapted from Refs. \cite{Theers2016} and \cite{Theers2018}, respectively, under a CC BY 3.0 license. Published by The Royal Society of Chemistry. (C) (i) Janus motor with catalytic cap that converts MPCD fluid species A into species B. The resulting diffusiophoretic force propels the motor. Cap size is determined by angle $\theta_{\mathrm{C}}$. (ii) Concentration and flow field near a Janus motor. (iii) Collective behavior of motors as a function of $\theta_{\mathrm{C}}$ and volume fraction $\phi$. Adapted from Ref.\ \cite{Huang2017} under a CC BY 3.0 license. Flow fields are shown in the laboratory frame. \label{fig:3}}
\end{figure}

 Stark and coworkers \cite{Alizadehrad2015} used MPCD simulations to investigate the swimming motions of the protozoan parasite \textit{Trypanosoma brucei}. The trypanosome swims by beating an attached flagellum that is wrapped around its spindle-shaped body in a left-handed helical half-turn (Fig.\ \ref{fig:3}(A)) \cite{Alizadehrad2015, Heddergott2012}.  The attachment of the flagellum distorts the trypanosome's flexible cell body into an asymmetric chiral shape, which results in corkscrew-like swimming trajectories that are asymmetric under time reversal. The cell body and flagellum were modeled as an elastic network of small beads connected via harmonic springs. A time-dependent and spatially varying potential was applied along the flagellum to drive its sinusoidal beating motion. The model accurately described the trypanosome's corkscrew-like forward swimming mode, and it was found that the location of the flagellum along the trypanosome body maximizes its swimming speed. Similar MPCD-based models have been used to elucidate features for swimming of other unicellular flagellates including sperm cells \cite{Rode2018} and the bacterium \textit{Escherichia coli} \cite{Hu2015} (Fig.\ \ref{fig:3}(A)).

\par MPCD simulations have also been used to investigate collective behaviors of active solutes such as clustering and gas--liquid phase separation. In passive systems, these phenomena reflect metastable or equilibrium phases arising from attractive interactions between particles. In active systems, by contrast, they are nonequilibrium steady states that are thought to arise from particle motility, steric repulsions, and HI \cite{Cates2015, Theers2018}.  Understanding motility-induced clustering (MIC) and motility-induced phase separation (MIPS) in active systems is key to controlling processes such as the formation of the surface-attached bacterial aggregates responsible for biofouling and to develop new routes to self-assemble particles into novel materials \cite{Cates2015}.

\par Theers et al.\ \cite{Theers2018} used MPCD to study the effect of particle shape and hydrodynamics on the collective behavior of rigid, prolate spheroidal squirmers in the quasi-two-dimensional confinement of a narrow slit (Fig.\ \ref{fig:3}(B)). Squirmers are generic models for self-propelled solutes. They differ in the type of active stress applied to the surrounding fluid to mimic the flow fields generated by the appendages (flagellum or cilia) of microswimmer classes, including pushers (e.g., \textit{Escherichia coli}), pullers (e.g., \textit{Chlamydomonas}), and neutral swimmers (e.g., \textit{Paramecium}) (Fig.\ \ref{fig:3}(B)). In the simulations, the swimming velocity and active stress were specified by imposing a spatially varying slip velocity across the squirmer surface. Comparison with complementary simulations of non-hydrodynamic active Brownian particles showed that HI suppress and enhance MIPS in systems of neutral spherical and elongated squirmers, respectively, indicating a remarkable sensitivity to swimmer shape (Fig.\ \ref{fig:3}(B)).  Further, whereas pullers exhibited an increased tendency to phase separate over neutral swimmers, MIPS was suppressed in systems of pushers. Hence, these findings reveal that collective behavior of active solutes can depend on the complex interplay between HI, particle geometry, and propulsion mechanism.

\par  The collective behavior of self-propelled chemical motors has also been simulated using the MPCD framework. Kapral and coworkers \cite{Huang2017} developed a model for Janus motors, which are spherical colloids with asymmetric catalytic activity on their surface (Fig.\ \ref{fig:3}(C)). Reactions occurring on the catalytically active side of the surface generate a local concentration gradient of chemical species and a diffusiophoretic force that propels the particle. The MPCD solvent was modeled as a binary A--B mixture of solvent species that interact with the Janus particle through hard reflective collisions during the MPCD streaming step. The two species were identical, but component A was given a slightly larger collision radius. Particles of species A that collided with the catalytic cap (C) were converted to species B to model the irreversible reaction $\rm{A} + \rm{C} \rightarrow \rm{B} + \rm{C}$.  Steady-state conditions were maintained by simulating the reaction $ \rm{B} \rightarrow {A}$ in the bulk using a reactive MPCD scheme \cite{Rohlf2008}.  Janus motors with small catalytic caps were found to exhibit an increased propensity to cluster at high particle volume fractions (Fig.\ \ref{fig:3}(C)). Particles with large caps, by contrast, tended to exhibit MIC at low densities.  This behavior suggests that catalytic cap size and particle volume fraction are key design parameters in controlling the self-assembly and collective dynamics of Janus motors.

\section{Conclusions and outlook} \label{S:4}
Over the past twenty years, MPCD has emerged as a powerful mesoscale technique for simulating HI in soft matter systems such as flowing complex fluids, sedimenting colloidal suspensions, swimming microorganisms, and self-propelling chemical motors. Massively parallel implementations of the algorithm in open-source molecular simulation packages have enabled larger, longer MPCD simulations and lowered the barrier to application of the method. Looking forward, the MPCD method may have promising applications in adaptive resolution schemes where the level of coarse-graining varies in space \cite{praprotnik:jcp:2005}, or in hybrid particle-continuum simulations that utilize microscopic simulations for generating on-the-fly constitutive models for continuum computational fluid dynamics \cite{stalter:cpc:2018}.

Despite extensive prior research, there still remain several aspects of the MPCD method in need of further investigation. First, it is challenging to directly map the transport coefficients of the MPCD solvent onto a real liquid (Section \ref{S:2:mapping}), partially due to the thermal fluctuations in the solvent that are inherent to the collision rule and are hence difficult to tune. The result is a fluid with a high self-diffusivity and low Schmidt number; there is an open question as to whether a solvent with these properties adequately reproduces the relevant HI in the Stokes flow regime \cite{shakeri:pf:2018}. Second, it has been suggested that the anisotropy of the cells underlying the MPCD collision may introduce undesirable artifacts. The recently proposed isotropic SRD collision rule \cite{muhlbauer:prf:2017} circumvents this issue, but additional work is needed to understand the strengths and limitations of such schemes. Finally, MPCD has had only limited application to multiphase fluids \cite{gompper:advcomp:2009, eisenstecken:epl:2018}, a technologically important class of complex fluids. Future advances in these areas will  greatly expand the capabilities of MPCD and solidify its status as one of the most flexible and robust methods for modeling HI in soft materials.

\section{Conflicts of interest}
The authors declare no conflicts of interest. 

\section{Acknowledgements} 
We thank Jacinta Conrad for comments on the manuscript and discussions on bacterial motility. Work by M.P.H. was supported as part of the Center for Materials for Water and Energy Systems, an Energy Frontier Research Center funded by the U.S. Department of Energy, Office of Science, Basic Energy Sciences under Award \#DE-SC0019272. A.N. received support from the German Research Foundation (DFG) under project number NI 1487/2-1. J.C.P gratefully acknowledges support from the Welch Foundation (Grant E-1882) and the National Science Foundation (CBET-1705968).

\section{References and recommended reading} 
\bibliographystyle{model1-num-names}
\bibliography{mpcd_perspective}

\end{document}